\documentclass[epj ,nopacs]{svjour}
\usepackage{multirow}
\usepackage{adjustbox}
\usepackage{graphicx}
\usepackage{subfig}
\usepackage{rotating}
\usepackage{tabularx}
\usepackage{array, makecell, bigstrut}
\usepackage{threeparttable}
\usepackage{pdflscape}
\usepackage{afterpage}
\usepackage{mathtools}
\usepackage{orcidlink}
\usepackage{textgreek}
\begin{document}
\title{Development of MMC-based lithium molybdate cryogenic calorimeters for AMoRE-II 
}
\author{
A.~Agrawal\,\orcidlink{0000-0001-7740-5637},
V.V.~Alenkov\,\orcidlink{0009-0008-8839-0010},
P.~Aryal\,\orcidlink{0000-0003-4955-6838},
H.~Bae\,\orcidlink{0000-0003-1393-8631},
J.~Beyer\,\orcidlink{0000-0001-9343-0728},
B.~Bhandari\,\orcidlink{0009-0009-7710-6202},
R.S.~Boiko\,\orcidlink{0000-0001-7017-8793},
K.~Boonin\,\orcidlink{0000-0003-4757-7926},
O.~Buzanov\,\orcidlink{0000-0002-7532-5710},
C.R.~Byeon\,\orcidlink{0009-0002-6567-5925}
% ...
}                     
%
%
%
\date{Received: date / Revised version: date}
%
\abstract{
The AMoRE collaboration searches for neutrinoless double beta decay of $^{100}$Mo using
molybdate scintillating crystals via low temperature thermal calorimetric detection. The early phases of the experiment, AMoRE-pilot and AMoRE-I, have demonstrated competitive discovery potential. Presently, the AMoRE-II experiment, featuring a large detector array with about 90 kg of $^{100}$Mo isotope, is under construction.  This paper discusses the baseline design and characterization of the lithium molybdate cryogenic calorimeters to be used in the AMoRE-II detector modules. 
The results from prototype setups that incorporate new housing structures and two different crystal masses (316 g and 517 -- 521 g), operated at 10 mK temperature, show energy resolutions (FWHM) of 7.55 -- 8.82 keV at the 2.615 MeV $^{208}$Tl $\gamma$ line and effective light detection of 0.79 -- 0.96 keV/MeV. The simultaneous heat and light detection enables clear separation of alpha particles
with a discrimination power of 12.37 -- 19.50 at the energy region around $^{6}$Li$(n,\alpha)^3$H with Q-value = 4.785 MeV.  
Promising detector performances were demonstrated at temperatures as high as 30 mK, which relaxes the temperature constraints for operating the large AMoRE-II array.  
\PACS{
      {PACS-key}{discribing text of that key}   \and
      {PACS-key}{discribing text of that key}
     } 
} 
\maketitle
\section{Introduction}
\label{intro}
Neutrinos are the only fermions with no electric charge and have been observed with single-handedness (only left-handed neutrinos or right-handed antineutrinos). 
The existence of neutrino oscillation demonstrates that neutrinos have mass, even if it is very tiny~\cite{nu_osc}. 
One simple scenario to explain these findings may be lightly massive neutrinos with a Majorana nature, in which the neutrino particles are the same as the anti-neutrino particles.  
The existence of neutrinoless double beta decay (0$\nu$DBD) would probe the Majorana nature of neutrinos since
it involves a lepton-number violating transition between a neutrino and an antineutrino~\cite{dbd_blackbox}.

For certain even-even nuclei, which are more tightly bound than their odd-odd neighbors with the same mass number, double beta decay (DBD) is the primary decay channel. There exist 35 candidate nuclides that meet this condition. In the Standard Model of particle physics, DBD with two neutrino emissions ($2\nu$DBD) is possible and  has been observed for 11 candidate nuclides with half-lives ($T_{1/2}$) that range from 10$^{18}$ to 10$^{24}$ years~\cite{2nbb}. 
As with single beta decay, neutrinos carry away energy from $2\nu$DBD, resulting in observed energy spectra with broad energy distributions. Since $0\nu$DBD does not emit neutrinos in the final product, its presence can be confirmed by an excess of event counts at the endpoint of the two-electron energy spectrum, where the energy equals the Q-value.
Since 0$\nu$DBD does not conserve lepton number,  the discovery of the 0$\nu$DBD will have far-reaching implications in understanding the leptogenesis scenarios
and the baryon asymmetry of the Universe.  Over the past years, the importance of demonstrating the existence of $0\nu$DBD has been recognized and motivated ambitious experimental efforts worldwide~\cite{0nbb_review}.

The AMoRE experiment is searching for 0$\nu$DBD of the $^{100}$Mo nuclide. The Q-value of this decay is 3.034 MeV, so the energy region of interest (ROI) for the signal events is higher than 2.6 MeV, which represents the limit 
above which
gamma background from naturally occurring nuclides are low. 
$^{100}$Mo has a natural abundance of 9.7\%, which is relatively high among the candidate nuclides for DBD.
Mo samples to be used in experiments for $0\nu$DBD can be highly enriched with $^{100}$Mo by centrifuges, enabling a large target exposure. 
In the AMoRE experiment, large scintillating molybdate crystals grown from molybdate powder with about 96\%  $^{100}$Mo enrichment are used.
The temperature increase and the light emission produced by the energy deposition from particle interaction are detected by very precise sensors based on metallic magnetic calorimeters (MMCs). 
Dual detection with the MMC sensors provides the capability for distinguishing
between $\gamma$, $\beta$ induced events from those caused by $\alpha$ decay~\cite{gbkim_2017,amore_det_principle,wtkim_jinst2022}.

The AMoRE collaboration has already demonstrated the feasibility of this approach
in the early phases of the experiment.
AMoRE-pilot, the proof-of-concept phase built with six calcium molybdate crystal detectors with a total crystal mass of 1.87 kg, was successfully operated at Yangyang underground laboratory (Y2L) from 2016 to 2018~\cite{amore_pilot_2019,amore_pilot_2024}. AMoRE-I, the successive phase with a goal of achieving the most stringent limit for $^{100}$Mo $0\nu$DBD, comprised thirteen calcium molybdate crystals and five lithium molybdate crystals with a combined mass of 6.19 kg, operated at Y2L between 2021 and 2023~\cite{hbkim_jltp2022,yoomin_amoreI2024}. These two early phases of AMoRE demonstrated promising discovery potential and provided guidance for the subsequent large-scale experiment, AMoRE-II, that will ultimately consist of around 360 molybdate crystals with a total crystal mass of about 157 kg corresponding to approximately 90 kg of $^{100}$Mo~\cite{sckim_nc2023,ymoh_neutrino2024}. 
The experiment aims to reach a sensitivity of $T_{1/2}$ = $4.5 \times 10^{26}$ years for 0$\nu$DBD of $^{100}$Mo. This corresponds to an effective Majorana neutrino mass of 18 -- 31 meV, depending on the nuclear matrix elements~\cite{nme_mo100_1,nme_mo100_2}, under the assumption of light Majorana neutrino exchange.
In the preparations for AMoRE-II, 
considerable efforts have been directed toward optimizing detector performance and
standardizing fabrication steps with sufficient quality control to achieve a reliable production of the large array of crystal detectors. 
This paper elaborates on the development of the AMoRE-II detector modules and presents the outcome of their performance tests. 

\section{Experimental setup}
\label{sec:1}
\subsection{Detector principle}
\label{sec:2}
Low-temperature detectors that are used for thermal calorimetric detection typically operate at temperatures of a few tens of millikelvin.
At these temperatures, the heat capacity of the crystal absorber with a mass of a few hundred grams is low enough ($\sim$1 nJ/K) to support high resolution measurements of energy depositions of a few MeV. 
When a particle deposits energy in a dielectric crystal, highly energetic phonons 
close to Debye frequencies are generated. These phonons quickly decay via
anharmonic processes to lower-energy phonons at a rate proportional to the fifth power of their energy. Once they reach a state where they travel ballistically
with a mean free path longer than the dimension of the typical crystal absorber
(a few centimeters), these phonons, usually called athermal phonons, travel around the crystal on a timescale of a  few milliseconds before thermalization~\cite{probst,sust_review}.
The phonon thermalization process, which results in a thermal phonon distribution
at a specific temperature, includes isotope and defect scattering in the bulk
and inelastic surface scatterings. 
By introducing a small metal film evaporated on the crystal surface, athermal phonons
and their thermalization can be detected via the film, thereby serving as a phonon 
collector.
In the AMoRE setups, an MMC that is thermally connected to the phonon collector detects the corresponding temperature change, which is reflected as the change in magnetization of the paramagnetic sensor. 
The athermal and thermal processes of heat flow among the detector components
are described in detail in references~\cite{epjplus_2023,jltp_thermal_model_2024}.
The MMC is equipped with a superconducting current loop with persistent current. The change in magnetization introduces a screening current in the superconducting current loop that is measured by a SQUID, leading to a voltage signal that can be recorded as the heat channel signal~\cite{amore_det_principle}. 

In the case of a scintillating crystal, a small fraction (a few percent) of the absorbed energy generates scintillation photons, while
most of the energy converts into phonons in the crystal lattice. 
An addtional detector setup with a wafer of Si or Ge and an MMC sensor
is used to detect the scintillation photons in 
the AMoRE crystals (light channel)~\cite{hjlee_LD2015,mbkim_ieee2023}. 
The early detector development of the heat and light detection is described in previous reports~\cite{gbkim_2017,amore_det_principle,wtkim_jinst2022,wtkim_jltp2022}.

\subsection{AMoRE-II detector}

\subsubsection{Baseline design of the detector module}\label{sec:amore2_det}
\begin{figure}
\centering
\subfloat[][\centering ]{\includegraphics[width=1\linewidth]{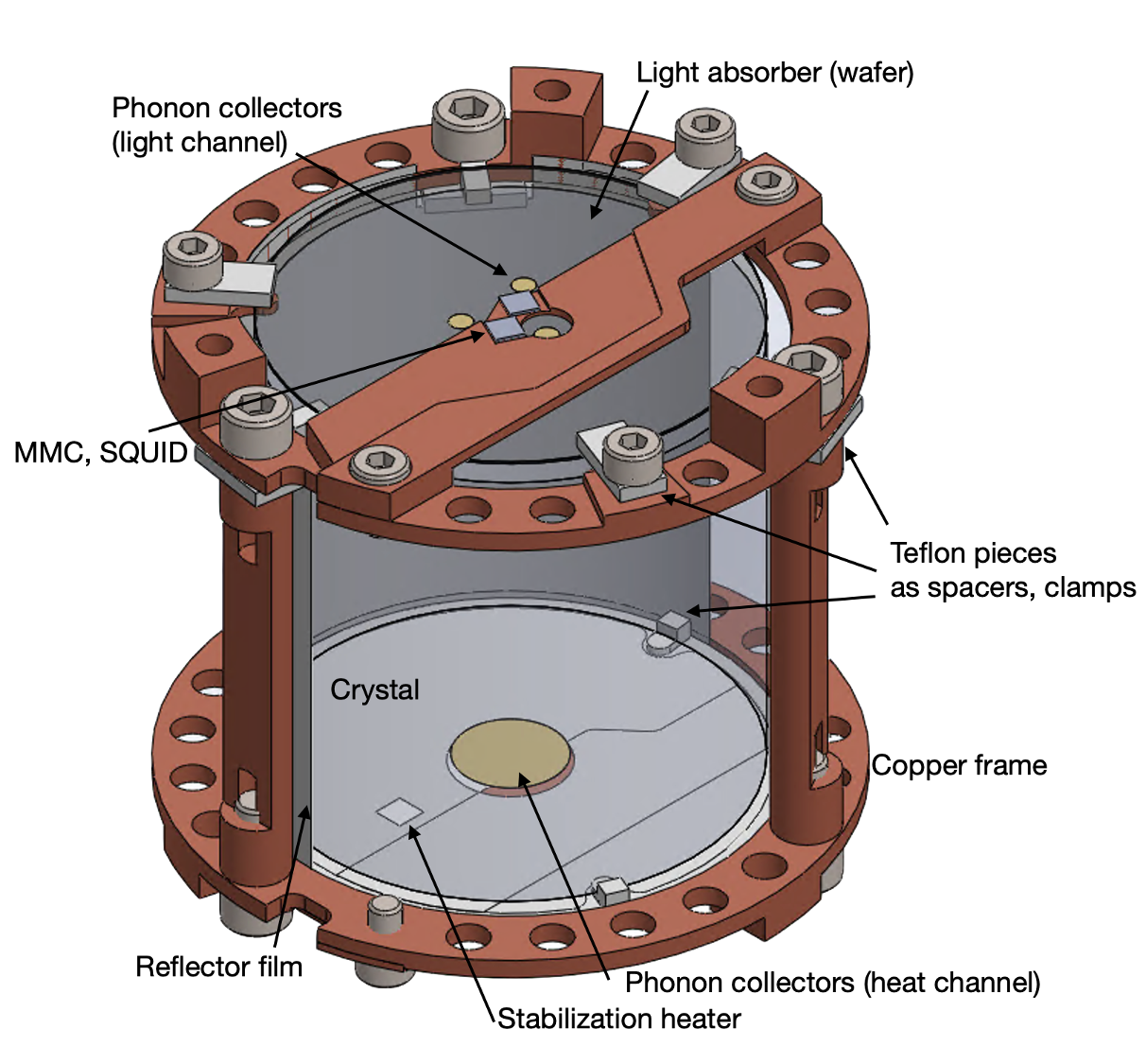}}\\
\subfloat[][\centering ]{\includegraphics[width=0.46\linewidth]{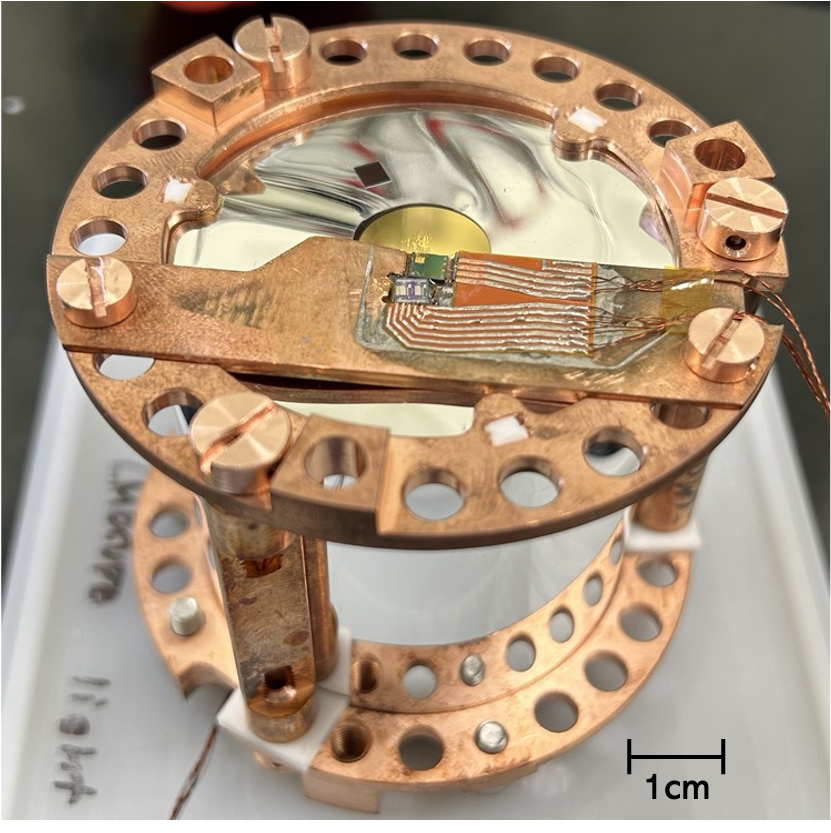}}
\quad
\subfloat[][\centering ]{\includegraphics[width=0.43\linewidth]{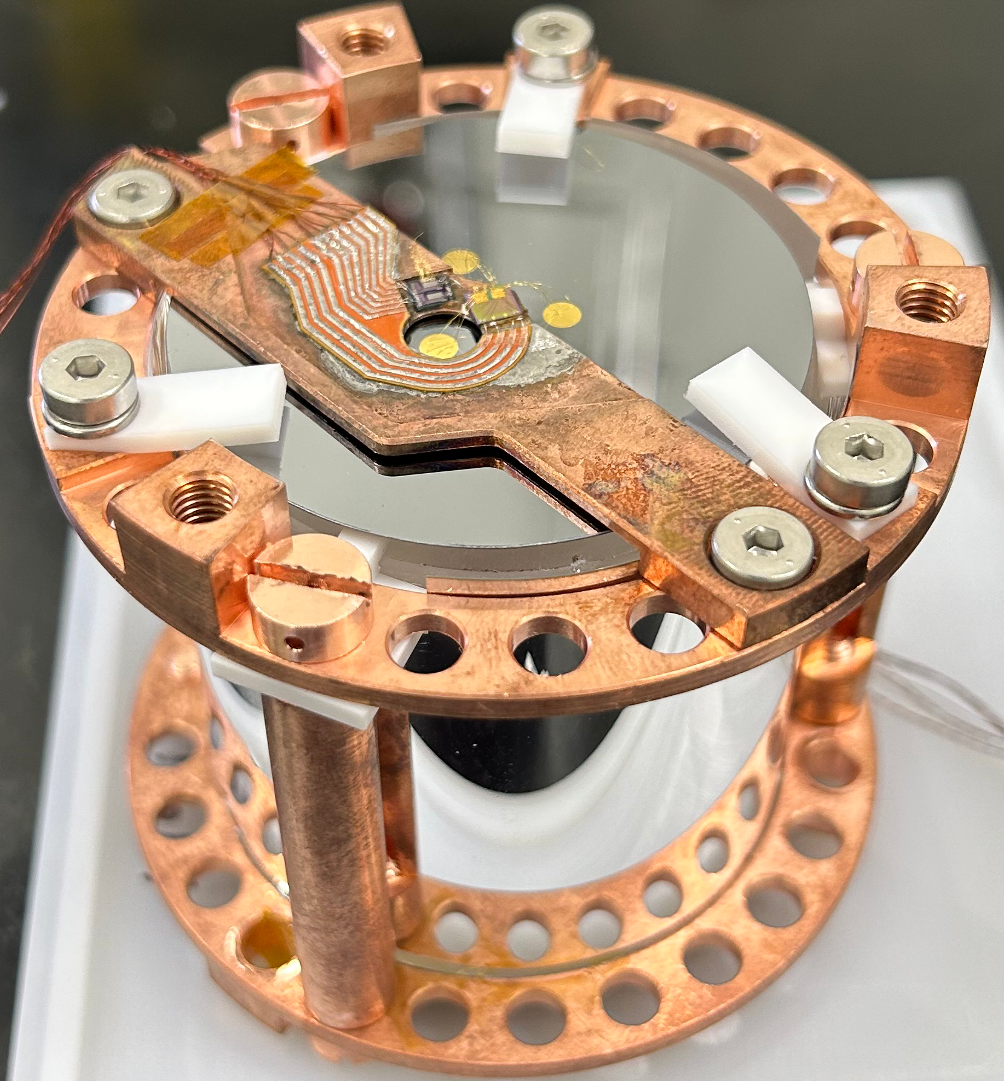}}
\caption{The AMoRE-II detector module.
(a) Drawing of the module.
Photos of an example R\&D setup (DET2, see Tab.~\ref{tab:dets}) showing the heat channel side (b) and the light channel side (c).}
\label{fig:1}       
\end{figure}
Based on previous phases of the experiments and R\&D studies, we have developed the baseline design  of the detector module for the AMoRE-II experiment. The crystal absorbers will
be $^{100}$Mo enriched lithium molybdate (Li$_2$$^{100}$MoO$_4$) crystals. 
Cylindrical crystals with two different dimensions will be used: 5 cm (Height) $\times$ 5 cm (Diameter) and 6 cm (H) $\times$ 6 cm (D). The masses of the two sizes are about 300 g and 520 g, respectively. 
Since we recently found that the crystals with diffusive surfaces show less position-dependent behavior and better energy resolution~\cite{wtkim_jinst2022}, we decided that a significant fraction of the newly produced crystals will have diffusive rather than polished surfaces.
For the phonon collector of the heat channel, we use a gold film with a molybdenum adhesion layer. The diameter of the phonon collector is 14 mm, and the thickness is 300 nm for the gold layer and 5 nm for the molybdenum adhesion layer. 
The AMoRE-II MMC sensor is a Ag:Er film on a meander-shaped superconducting loop~\cite{mmc}. Two-stage 
DC SQUIDs are adopted to measure the induced current in the meander of the MMC originating from particle interaction~\cite{squid_ptb}. A thermal connection
between the MMC and the phonon collector film is provided by 25 $\mu$m
diameter annealed gold wires. 

The holder structure of the detector module is made out of NOSV (Aurubis) copper with OFE copper (Arubis) or brass bolts and PTFE spacers and clamps~\cite{radioassay_frontier2024}.  
The module design uses the small amount of copper in order to reduce
its contribution to the overall weight of the detector array and potential radioactive background.
The mass of the copper frames is approximately 130 g for the 5 cm crystals and 150 g
for the 6 cm crystals. 
Additionally, compared to the previous light detectors, the gap between the Si wafer of the light detector and the crystal was reduced to 2.7 mm. It also allows the better enclosure by the reflector film.
To reduce the vibration noise, the light absorber is secured more tightly with PTFE clamps.

The light absorbers are silicon wafers with two dimensions in diameter and thickness (T): 5 cm (D) $\times$ 280 $\mu$m (T) for the 5 cm crystals and 6 cm (D) $\times$ 350 $\mu$m (T) for the 6 cm crystals. The wafers have a 90 nm thick SiO$_2$ coating on the side facing the crystal absorber to enhance the scintillation light absorption at the wafer surface.
The phonon collector for the light channel is patterned as three identical circles, each with a diameter of 3 -- 6 millimeters, composed of a 5 nm of Mo layer and a 300 nm of Au layer as the heat channel case.
The crystal is enclosed by a Vikuiti (3M) reflector film except for the side facing the light detector for an efficient light collection.

A stabilization heater (AuPd film on a Si wafer chip) will be implemented on the detector module to 
monitor the signal gain change due to the temperature drift.
 Figure~\ref{fig:1} shows the drawing of the AMoRE-II baseline design and photos of some R\&D setup built following the baseline design.

\subsubsection{Crystal preparation}
After an intensive survey of  scintillating molybdate crystals~\cite{hjkim_crystal_survey2019}, 
lithium molybdate (``LMO" hereafter) crystals were selected for their relatively easy growth, high concentration of molybdenum (56 wt\%) and low 
radioactive background level in the produced crystals. 
The LMO crystals at low temperatures also demonstrated excellent particle identification via scintillation light measurement~\cite{wtkim_jinst2022,hlkim_nima2020}.

To obtain crystals with sufficiently low radioactive contamination, we have developed purification procedures for the raw materials: molybdenum trioxide, MoO$_3$ and lithium carbonate, Li$_2$CO$_3$ powders. $^{100}$MoO$_3$ powder with about 96\% enrichment of $^{100}$Mo in the molybdenum element has been purchased from JCS ECP, Russia~\cite{jcs},  as the initial material and went
through purification procedures developed at the Center for Underground Physics (CUP), Institute for Basic Science (IBS) ~\cite{purification_MoO3_jinst2019,purification_MoO3_jrnc2017,purification_MoO3_jinst2020}.
The procedures were optimized to maximize the yield efficiency for the final pure product (about 93\%) and minimize the unrecoverable losses of $^{100}$Mo isotopes to below 1.5 \%. 
About 150 kg of $^{100}$MoO$_3$ powder was purified at CUP for the last 2.5 years with production rate of 5 kg of pure product per month
~\cite{mo_powder_purification}. 
Lithium carbonate powder was obtained from two different sources. One is the old stock (NRMP TU 6-09-3728-83) preserved at the Nikolaev Institute
of Inorganic Chemistry (NIIC), and the other is powder purified at CUP using sorption, co-precipitation and carbonization methods~\cite{li_powder_purification_frontier2024}.  
Before the crystal growth, the purity of the precursors was verified
with ICP-MS and HPGe, exhibiting sufficiently low level of radioactive contamination~\cite{mo_powder_purification,li_powder_purification_frontier2024,li_powder_purification}.

The crystals were grown using the conventional Czochr\-alski (Cz) method at CUP~\cite{cup_lmo_crystal} and the low-thermal gradient Czochralski (LTG-Cz) method at NIIC~\cite{lmo_niic}. 
The conventional Cz method allows for high segregation of impurities, resulting in crystals with low radioactive contamination. However, this method has a low crystallization efficiency and large residual melts. In contrast, LTG-Cz shows high crystallization efficiency and is suitable for growing large crystals. The NIIC group demonstrated a capability for producing large-size LMO crystals that are larger than 7 cm (D) $\times$ 7 cm (H). However, as this process has a lower segregation factor, a follow-up double crystallization step is needed to satisfy the low radioactive background requirement.
The raw crystal boules from crystal growers are cut into cylindrical crystals in dimensions of 5 cm (D) $\times$ 5 cm (H) and 6 cm (D) $\times$ 6 cm (H). 
Some crystals have optical-quality polishing for all sides, while the remaining crystals have optical-quality polishing only on one base and diffusive surfaces elsewhere.
The phonon collectors are deposited on the polished base.

Thanks to continuous production at CUP and NIIC, the crystals with low radioactive contamination~\cite{radioassay_frontier2024} have been accumulated for several years, and currently we are at our peak production rate of 10 -- 15 crystals per month (5 -- 8 kg/month by crystal mass). 
As the AMoRE-II detectors will be built in two stages, the production will continue until 2025.  Since the LMO crystals are very vulnerable to air's moisture damage, 
we established a dedicated protocol for reliable crystal storage.  After the final surface treatment, mineral oil is applied to all the surfaces of the crystals.
Afterward, each crystal is packed with three layers of vacuum-sealed aluminized mylar bags. The packages containing the crystals are stored in desiccators with constant nitrogen gas flow.

\subsubsection{Detector fabrication}

The crystal surface must be thoroughly cleaned to ensure proper adhesion of the phonon collector on the crystals.
It is crucial to avoid moisture and water damage during the crystal cleaning because LMO crystals are very hygroscopic. 
We first use kerosene with sonication to remove
the oil layers which was applied for the crystal storage.
Afterward, further cleaning procedures are conducted, including sonication cleaning and manual swiping in  isopropyl alcohol (IPA) and acetone.
This process is repeated to ensure clean surfaces. After the solvent cleaning, an oxygen plasma ashing process removes any remaining organic residues.
Following the cleaning procedures, the phonon collectors are deposited on the crystals using e-beam evaporation.
For the light absorber, phonon collectors are similarly deposited on silicon wafers after adequate wafer cleaning. 
During the processes in the cleanroom, the crystal exposure to ambient moisture is unavoidable. Therefore, care should be taken to minimize exposure by controlling process times and humidity level. 

The crystals with phonon collectors are assembled into the detector holder alongside SQUIDs, MMCs and Vikuiti reflector
film. 
The SQUID and MMC sensors are glued to the Cu sensor plates utilizing an epoxy, STYCAST 1266 (Loctite) and
the polyimide printed circuit board (PCB), facilitating electrical connections for the sensor readouts, is attached to the Cu sensor plates using clean lead soldering~\cite{radioassay_frontier2024}. 
The detector assembly with the LMO crystals is carried out in a low-humidity room with water content less than 1000 ppm in the air. 
The electronic connections among SQUID, MMC and the PCB in the detector modules are made with Al wires with a diameter of 32 $\mu$m using a wedge wire bonding machine. 
The assembled detector modules are kept in a desiccator with a low humidity until they are moved to be installed at the cryostat.

\subsection{R\&D setup of the detector modules} 
We report about the characterization of four detector setups as AMoRE-II R\&D detector modules that were built based on the details discussed in subsection~\ref{sec:amore2_det}.
The specifications of each setup are described in Table~\ref{tab:dets}.  In the AMoRE-I experiment, MMCs with two meander-shaped pickup coils,
each covered by Ag:Er layers to form a gradiometric flux transformer, were utilized~\cite{mmc_meander,mmc_meander_kriss}. In the R\&D experiments, we tested MMCs where only one pickup coil
is covered by the Ag:Er layer to investigate the temper\-ature-dependent baseline change to establish the magnetization as a function of temperature.
Although MMCs with two Ag:Er areas (``double side MMC") are considered better for cancelling noises common to both areas, we found that MMCs with one Ag:Er area (``one side MMC") provide a viable solution. The DC baseline in the one side MMCs allows for correction of gain changes due to temperature drift and does not suffer from serious noise issue.
In the detector setups discussed in this report, except for one light detector, which was paired in DET2, one side MMCs were used.
We previously reported the reduced noise level and improved energy resolution using MMCs with no thermal link and followed this approach for one detector (DET3) as specified in Table~\ref{tab:dets}~\cite{wtkim_jinst2022}. 
There were attempts to improve the performance of the light channel, including optimization of the phonon collector (annealing and increased area) and thermal conductance between the phonon collector and the MMC~\cite{mbkim_ieee2023}. These efforts were reflected in some detector modules such as DET3 and DET4.  
For the R\&D setups, normal copper frames were also used in addition to NOSV copper frames. A strict protocol for the cleanliness of materials was not applied since the test was conducted at an above-ground cryostat with high environmental radioactive background and the primary focus was on the detector performance test.  

\begin{figure}
\centering
\subfloat[][\centering ]{\includegraphics[width=0.7\linewidth]{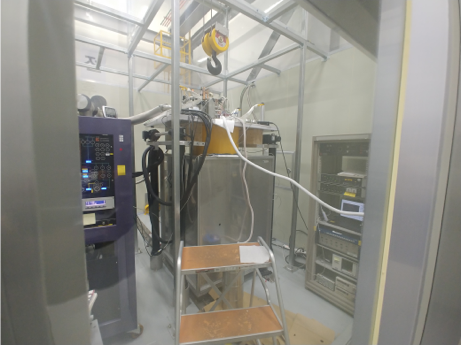}}\\
\subfloat[][\centering ]{\includegraphics[width=0.7\linewidth, angle=0, origin=c]{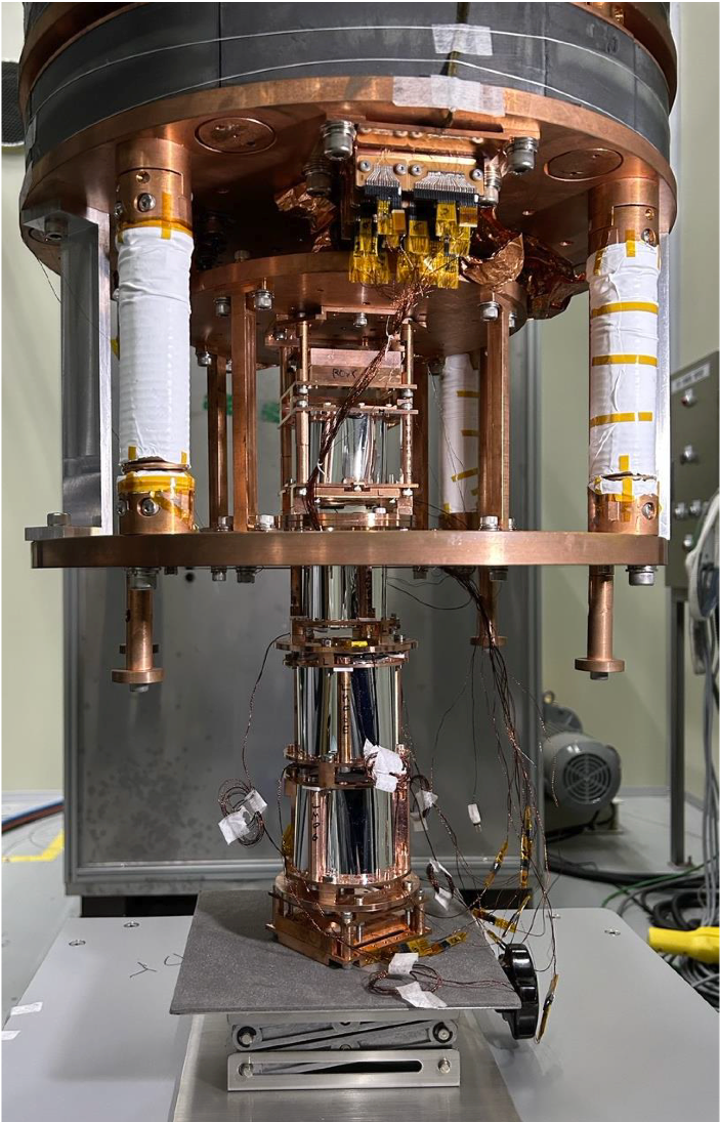}}
\caption{Cryostat system for the R\&D experiment. (a) The cryostat in the low humidity facility. 
(b) A detector tower with four detector setups, mounted under the vibration damping system of the cryostat.}\label{rody_photo}
\end{figure}

We used the above-ground cryogen-free dilution refrigerator, which is located in CUP, IBS headquarter in Daejeon, South Korea. It is a large cryostat that accommodates a 10 cm thick lead layer, which shields radiation from the top direction, and a vibration damping structure~\cite{vibration_damper} under which the detector tower of multiple detectors is installed at the mixing chamber stage. The cryostat is enveloped by 10 cm thick layers of lead bricks along the sides and underneath the bottom to substantially reduce environmental radiation background. The base
temperature is below 10 mK and the cooling power is a few $\mu$W at 10 mK. We tested the performance of the detectors at temperatures of 10, 20 and 30 mK at the mixing chamber stage. 
To avoid crystal damage caused by exposure to ambient moisture,
the cryostat is located in the low-humidity facility, which provides a humidity level below 10 \% RH at room temperature when we install the detectors. 
Figure~\ref{rody_photo} shows the low-humidity facility with the cryostat and a detector tower installed at the mixing chamber stage of the cryostat with the vibration damper. 

The detector signals were saved continuously using a multi-channel ADC digitizer. In digitization, the sampling rate was set as 100k samples/s with 18-bit resolution and a dynamic range of 10 V.
The ADC was optically connected to a trigger condition board (TCB), which controls the global clock and transfers data to a data storage computer, to minimize  noise interference in the signals. 
The event triggering was conducted offline with bandpass filtering.

\begin{table*}[t]
\centering
\begin{threeparttable}
\caption{Summary of AMoRE-II R\&D detector setups. All setups were built with  LMO crystals and Si wafer light absorbers.
All crystals have diffusive surfaces as described in \ref{sec:amore2_det}. }
\label{tab:dets}       
\begin{tabular}{>{\centering\arraybackslash}m{1.5cm}
>{\centering\arraybackslash}m{3.1cm}
>{\centering\arraybackslash}m{3.1cm}
>{\centering\arraybackslash}m{2.5cm}
>{\centering\arraybackslash}m{2.3cm}
>{\centering\arraybackslash}m{2.6cm}} 
\hline\noalign{\smallskip}
Detector Name & Crystal\tnote{a} \par (size\tnote{b}, mass, manufacturer)  & Light absorber size\tnote{c} & Phonon collector size on the light absorber\tnote{d} & Note & Detector ID\tnote{g}\\
\noalign{\smallskip}\hline\noalign{\smallskip}
 DET1& Li$_2$$^{\mathrm{Nat}}$MoO$_4$  \par  ($6\times6$, 517.3g , NIIC)  & 5 cm $\times$ 280 $\mu$m & D = 3 mm &  & LMO6SL-Exp1.22\\
\noalign{\smallskip}
 DET2& Li$_2$$^{100}$MoO$_4$ \par ($5.1\times5$, 316g, CUP)  & 5 cm  $\times$ 280 $\mu$m & D = 3 mm & & LMOCUPEII-Exp2.22\\
\noalign{\smallskip}
 DET3& Li$_2$$^{100}$MoO$_4$ \par ($6\times6$, 520.8g, NIIC)  & 6 cm  $\times$ 350 $\mu$m & D = 6 mm & No thermal link\tnote{e}   LD by Ref~\cite{mbkim_ieee2023}\tnote{f}&LMO6E-Exp2.22\\
\noalign{\smallskip}
 DET4& Li$_2$$^{\mathrm{Nat}}$MoO$_4$ \par ($6\times6$, 517.3g, NIIC)  & 6 cm  $\times$ 350 $\mu$m & D = 6 mm  & LD by Ref~\cite{mbkim_ieee2023}&LMO6-Exp2.22\\
\noalign{\smallskip}\hline
\end{tabular}
\begin{tablenotes}
       \item[a] Li$_2$$^{\mathrm{Nat}}$MoO$_4$ (natural Mo contents), Li$_2$$^{100}$MoO$_4$ ($^{100}$Mo enriched)
       \item[b] Cylindrical shape, diameter $\times$ height in cm 
       \item[c] Thin wafer, diameter $\times$ thickness
       \item[d] Patterned as 3 circles. Diameter of one circular pattern shown here
       \item[e] No direct wire for thermal connection between the MMC and the heat bath for the heat channel
       \item[f] Light detector built by the recipe in a reference~\cite{mbkim_ieee2023}
       \item[g] Detector ID for cross-referencing with other related reports
\end{tablenotes}
\end{threeparttable}
\end{table*}

\section{Results and Discussion}
\subsection{Pulse shape}
\begin{figure}
\centering
\subfloat[][\centering ]{\includegraphics[width=0.7\linewidth]{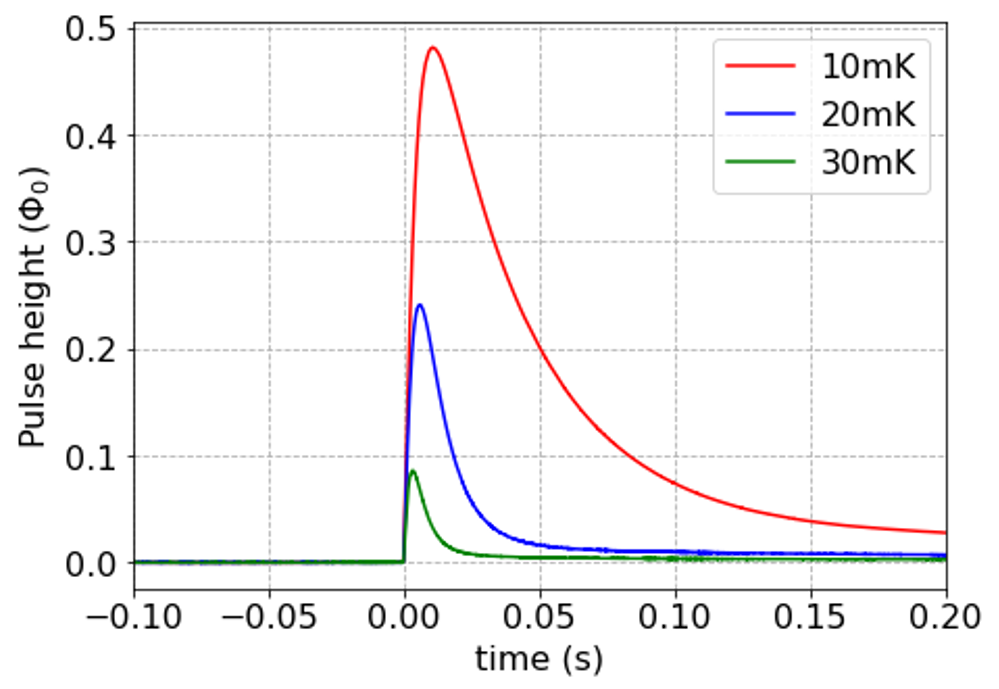}} \\
\subfloat[][\centering ]{\includegraphics[width=0.7\linewidth, angle=0, origin=c]{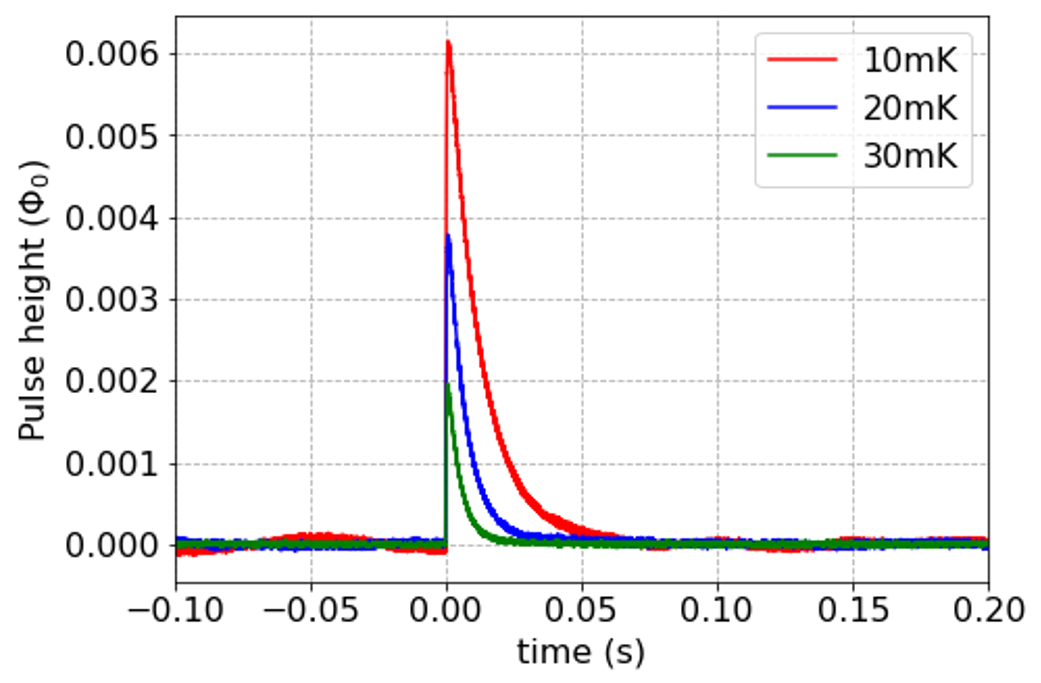}}
\caption{Signal pulses for the full absorption of 2.615 MeV gamma rays in a Li$_2$MoO$_4$ crystal at 10, 20, and 30 mK operation temperatures. The pulses were obtained by averaging many 
event signals from DET2. (a) Heat channel (phonon). (b) Light channel (scintillation). The pulse
height is represented as the flux in the SQUID in a unit of the magnetic flux quantum (\textPhi$_0$).}\label{signal_template}
\end{figure}

Fig.~\ref{signal_template} (a) shows the measured heat channel pulses for 2.615 MeV gamma rays ($^{208}$Tl) produced from a $^{232}$Th calibration source with DET2
at 10, 20, and 30 mK. The characteristic features of the pulses for overall detectors are summarized in Table~\ref{tab:pulses}.
The pulse shape for different detectors may depend on various factors, including the interface between the phonon collector and the crystal surface,
crystal conditions with different degrees of surface roughness, MMC response, the amount of MMC field current, and actual conductance of gold wires
which realize the thermal connections at MMC for the phonon collector and heat bath.
All the detectors showed reasonable signal responses up to 30 mK.  
The signal amplitude increase at lower temperatures, as expected, due to the lower heat capacities of the detector components and the temperature dependence of the sensor
responsivity~\cite{mmc}.
The rise times (the time elapsed between 10 and 90 \% of the pulse maximum on the rising edge) of
the heat channel signals range from 3.82 and 5.92 ms at 10 mK and become shorter as 1.22 -- 2.26 ms  at 30 mK.
Generally, a faster time response is beneficial for minimizing backgrounds in the ROI from unresolved pile-up events. 
Considering the signal and noise traces measured in the heat channel at 10 mK,
we estimated the rate of unresolved pile-ups from 2$\nu$DBD events to be less than
$3 \times 10^{-5}$ counts/keV/kg/year in the ROI~\cite{ymoh_neutrino2024}.
The rate is tolerable
within the AMoRE-II background budget, which aims for a level of less than 
$10^{-4}$ counts/keV/kg/year in the ROI. 
Futhermore, with the aid of fast light signals, the timing resolution
can be improved, and the rate of unresolved pile-ups can be suppressed. 
Alternatively, this rate can be further reduced at higher operational temperatures,
albeit with a trade-off in energy resolution due to smaller signal amplitudes.  
Regarding the decay time (the time between 90 and 10 \% of the height on the falling edge), it is generally in the range of 10 -- 40 ms
except for the signal of DET2 at 10 mK, which exhibits a decay time of about 111 ms, requiring further understanding.

\begin{figure}
\centering
\subfloat[][\centering ]{\includegraphics[width=0.7\linewidth]{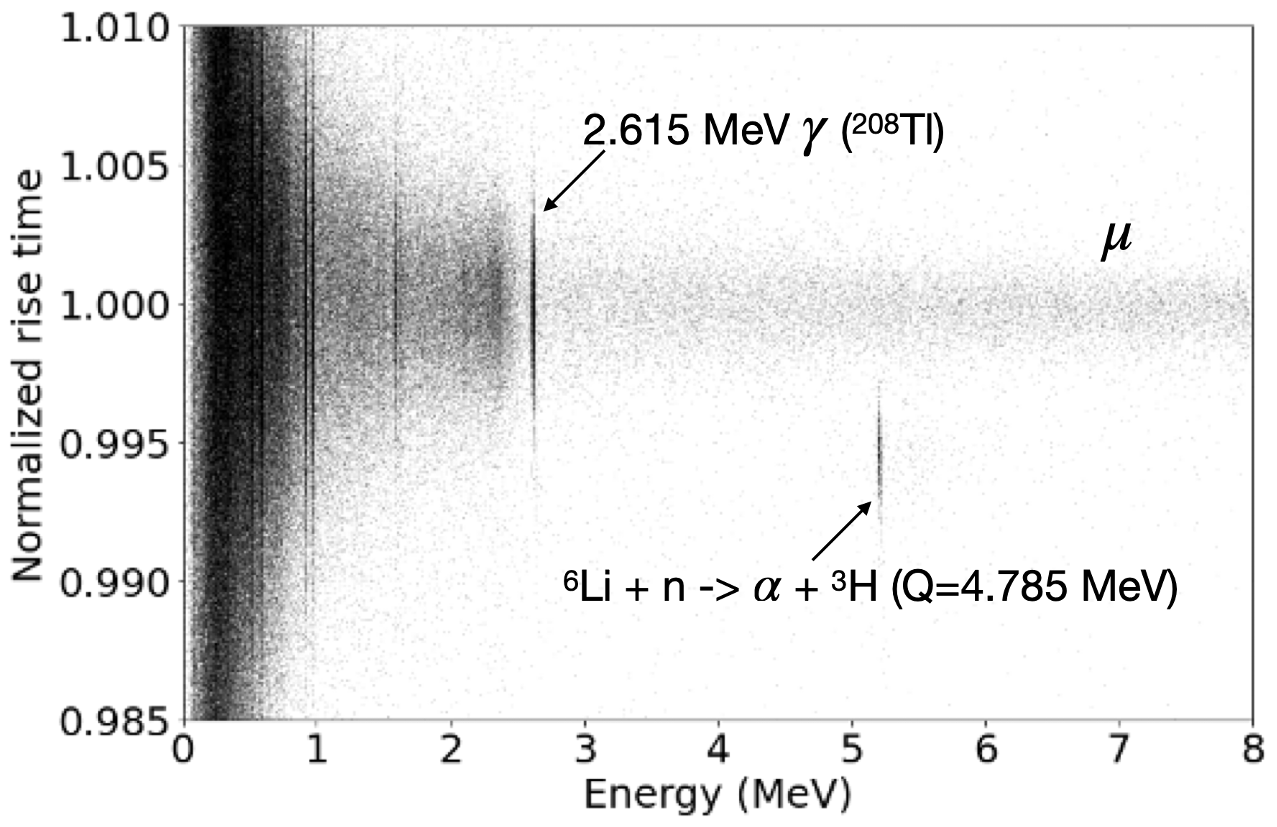}}\\
\subfloat[][\centering ]{\includegraphics[width=0.7\linewidth]{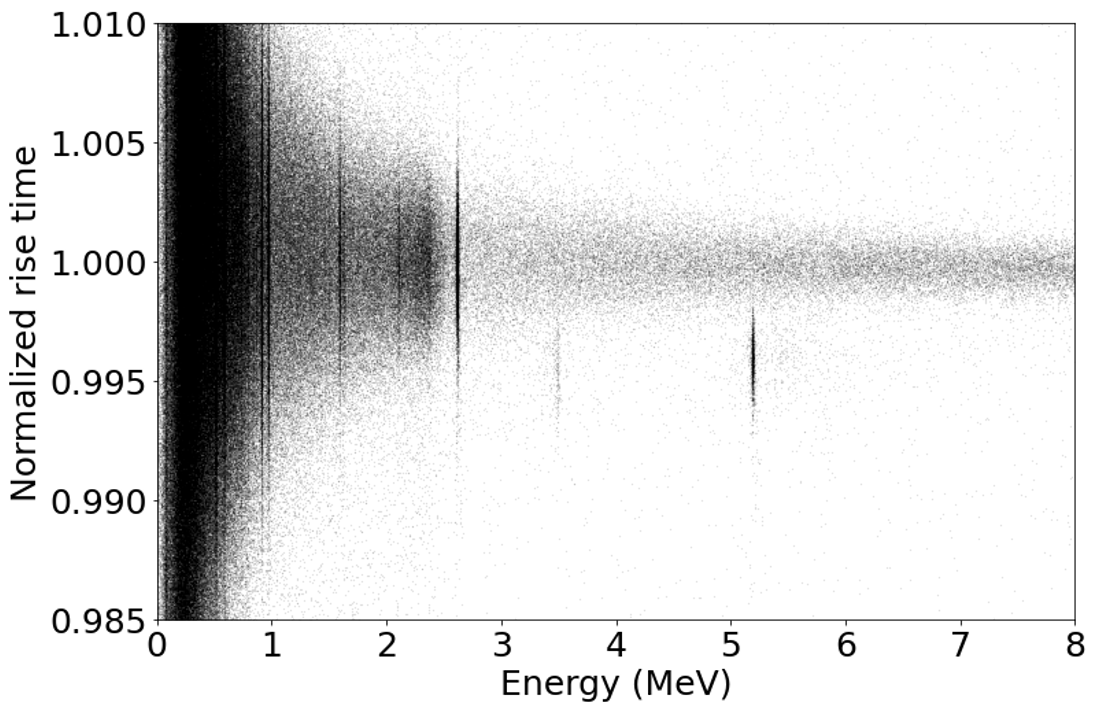}}\\
\subfloat[][\centering ]{\includegraphics[width=0.7\linewidth]{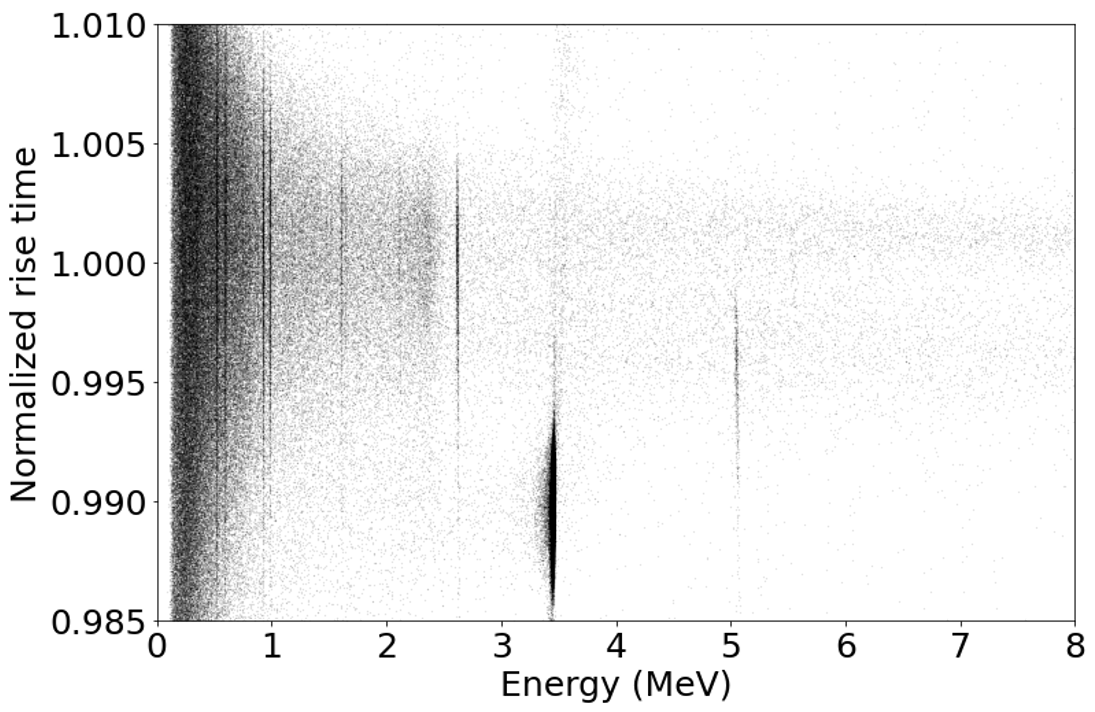}}\\
\subfloat[][\centering ]{\includegraphics[width=0.7\linewidth]{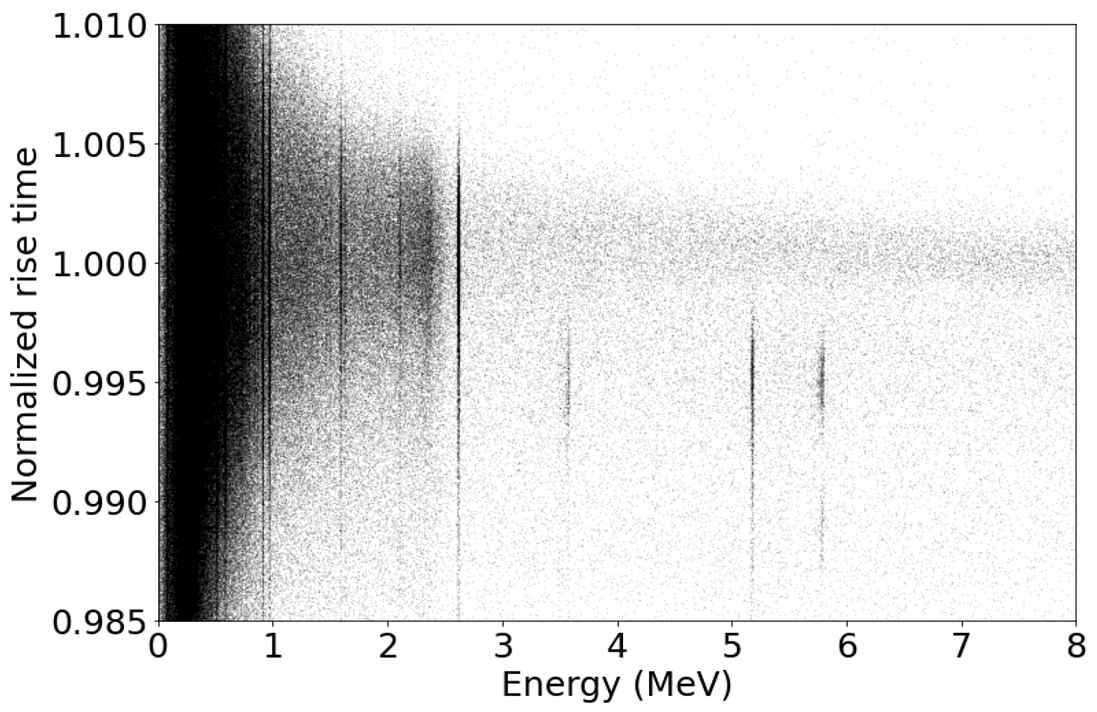}}
\caption{Rise time versus energy at 10 mK. Rise time is normalized to be one at the 2.615 MeV gamma line. 
(a) DET2. (b) DET3. (c) DET1. (d) DET4. 
Energy scales were calibrated with the gamma signals 
so that the alpha particle energies are slightly overestimated. 
The alpha peaks from $^{148}$Gd (Q = 3.271 MeV) and $^{210}$Po (Q = 5.407 MeV) sources are present for some detectors. 
}
\label{fig:RT_vs_E}      
\end{figure}

In previous experiments for crystals with polished surfaces,
a two-band
structure was clearly visible in a plot of energy versus rise time due to the muon
passage at the edge of the crystals~\cite{wtkim_jinst2022,wtkim_jltp2023}.
We reported that crystals with diffusive surfaces showed longer rise times of the signals and mitigation of position-dependent behavior~\cite{wtkim_jltp2023}. 
This effect resulted in a smaller spread in the rise time distribution and, therefore,
stronger pulse shape discrimination (PSD).
In the present setups prepared with diffusive crystal surfaces, this phenomena was confirmed even in a large crystal with dimensions of 6 cm (H) $\times$ 6 cm (D).
As shown in Fig.~\ref{fig:RT_vs_E} (a) and (b), the split structure is not apparent,
allowing clear discrimination of alpha events using pulse shape parameters such as rise time. 
The discrimination power (DP) is defined as:
\begin{equation}
\text{DP} = |m_{\beta/\gamma/\mu} - m_\alpha|/\sqrt{\sigma_{\beta/\gamma/\mu}^2 + \sigma_\alpha^2}
\end{equation}
where $m_{\beta/\gamma/\mu}$, $m_\alpha$ are the mean values of the parameters for non-alpha signals (beta, gamma, and muon signals) and alpha signals respectively, $\sigma_{\beta/\gamma/\mu}, \sigma_\alpha$ are the corresponding standard deviations. DP based on PSD between alpha particles and muons 
at the energy region around $^{6}$Li$(n,\alpha)^3$H (Q-value = 4.785 MeV)
are summarized in Table \ref{tab:performance}. 
The energy scales in Fig.~\ref{fig:RT_vs_E} are calibrated with distinct gamma lines, and
the energy of the alpha events is slightly overestimated due to different pulse shapes. 
At 10 mK, these detectors showed DP values larger than 3.
The R\&D test was not conducted long enough to collect a large amount of the degraded alpha (or surface alpha) events around the ROI, but the DP around 5 MeV, where we can find a good population of alpha events, can be a good indicator for the DP at the ROI.
DET1 and DET4, which we built with the exact same crystal, still showed a wide spread of rise times, as shown in Fig.~\ref{fig:RT_vs_E} (c) and (d).
We believe either its surface treatment was not optimal or the crystal might have a non-uniformity issue. 
A more detailed discussion about the position-dependent behavior of events with some detectors was presented in our earlier report~\cite{wtkim_jltp2023}.

\begin{table*}[t]
\centering
\caption{Summary of pulse shapes for the R\&D detectors at 10, 20, and 30 mK operation temperatures. The pulses were obtained by averaging many signals for 2.615 MeV gamma rays from $^{208}$Tl for both heat and light channels.
The pulse height is represented as the flux in the SQUID in a unit of the magnetic flux quantum (\textPhi$_0$).
The rise time is defined as the rise from 10 to 90\% of the pulse height on the rising edge
and the decay time as the fall from 90 to 10\% on the falling edge. The decay time of the light channel of DET1 may be incorrect by a few ms due to the slow vibration noise in the signal.}
\label{tab:pulses}       
\begin{tabular}{>{\centering\arraybackslash}m{2cm}
>{\centering\arraybackslash}m{1.8cm}
>{\centering\arraybackslash}m{2cm}
>{\centering\arraybackslash}m{2.4cm}
>{\centering\arraybackslash}m{0cm}
>{\centering\arraybackslash}m{2cm}
>{\centering\arraybackslash}m{2.5cm}
}
\hline\noalign{\smallskip}
\multirow{2}{*}{Detector name} & \multirow{2}{*}{\adjustbox{stack=cc}{Operation temperature (mK)}} &\multicolumn{2}{c}{ Heat channel} &&\multicolumn{2}{c}{ Light channel}\\
\cline{3-4} \cline{6-7}
& & Pulse height (\textPhi$_0$)  & Rise time / Decay time (ms) && Pulse height (m\textPhi$_0$)  & Rise time ($\mu$s) / Decay time (ms) \\
\noalign{\smallskip}\hline
\multirow{3}{*}{DET1} & 10 & 1.61 & 5.02 / 40.7  && 4.73 & 651 / 31.6\\
& 20 & 0.27 & 2.15 / 24.5  && 3.38 & 517 / 23.5 \\
& 30 & 0.05 & 1.38 / 33.3  && 2.09 & 468 / 26.6 \\
\hline
\multirow{3}{*}{DET2} & 10  & 0.48 &  5.82 / 110.7  && 6.20 & 461 / 28.2 \\
& 20 & 0.24  & 3.43  / 29.8 && 3.77 & 403 / 15.3\\
& 30 & 0.09 & 1.93  / 18.6 && 1.95 & 373 / 10.8 \\
\hline
\multirow{3}{*}{DET3} &10 &  0.29 & 5.11 / 31.8  && 4.06 & 671 / 18.9 \\
& 20 & 0.12 & 3.55 / 21.4 && 3.47 & 581 / 13.6\\
& 30 & 0.04 & 2.26 / 18.6 && 2.37 & 492 / 11.8\\
\hline
\multirow{3}{*}{DET4} & 10 & 1.39   & 3.82 / 28.5  && 4.52 & 873 / 134 \\
& 20 & 0.31 &  1.92 / 21.9 && 3.72 & 721 / 102\\
& 30 &  0.21 &  1.22 / 27.5 && 2.23 & 582 / 38.0\\
\hline
\end{tabular}

\end{table*}

\subsection{Energy resolution}
The experiments were conducted above ground with a $^{232}$Th source placed just outside the OVC of the cryostat to irradiate
the detector tower. The total event rate for each detector was in the range of  about 3 -- 4 Hz. 
Due to frequent high-energy events, including those from cosmic muons, and 
the typical recovery time for event signals to return to baseline being longer than 100 ms, there is a good chance that signal events appear in the long tail of
the preceding events. 
To determine the signal amplitudes, we applied a bandpass filter to the raw signals with passband frequencies chosen to minimize the influence of baseline fluctuations and other high-frequency noises~\cite{wtkim_nmo_2022}. 
For the heat channel signal, 
a first-order Butterworth bandpass filter was employed with typical lower bound frequencies ranging from
15 to 30 Hz and upper bound frequencies from 40 to 200 Hz.

\begin{figure}
\centering
\subfloat[][\centering ]{\includegraphics[width=0.47\linewidth]{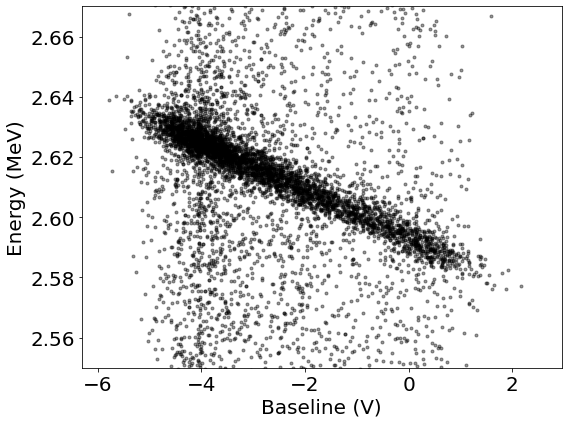}}
\quad
\subfloat[][\centering ]{\includegraphics[width=0.47\linewidth]{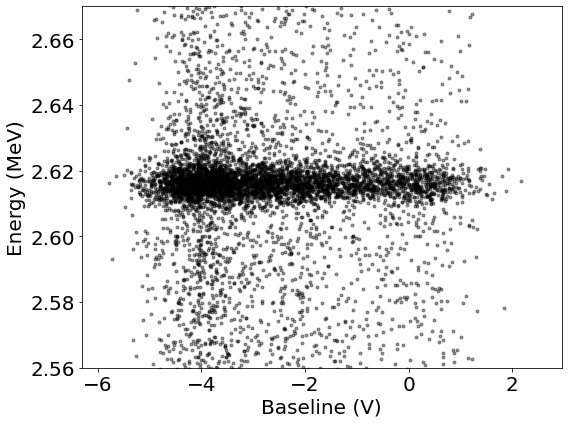}}\\
\subfloat[][\centering ]{\includegraphics[width=1\linewidth]{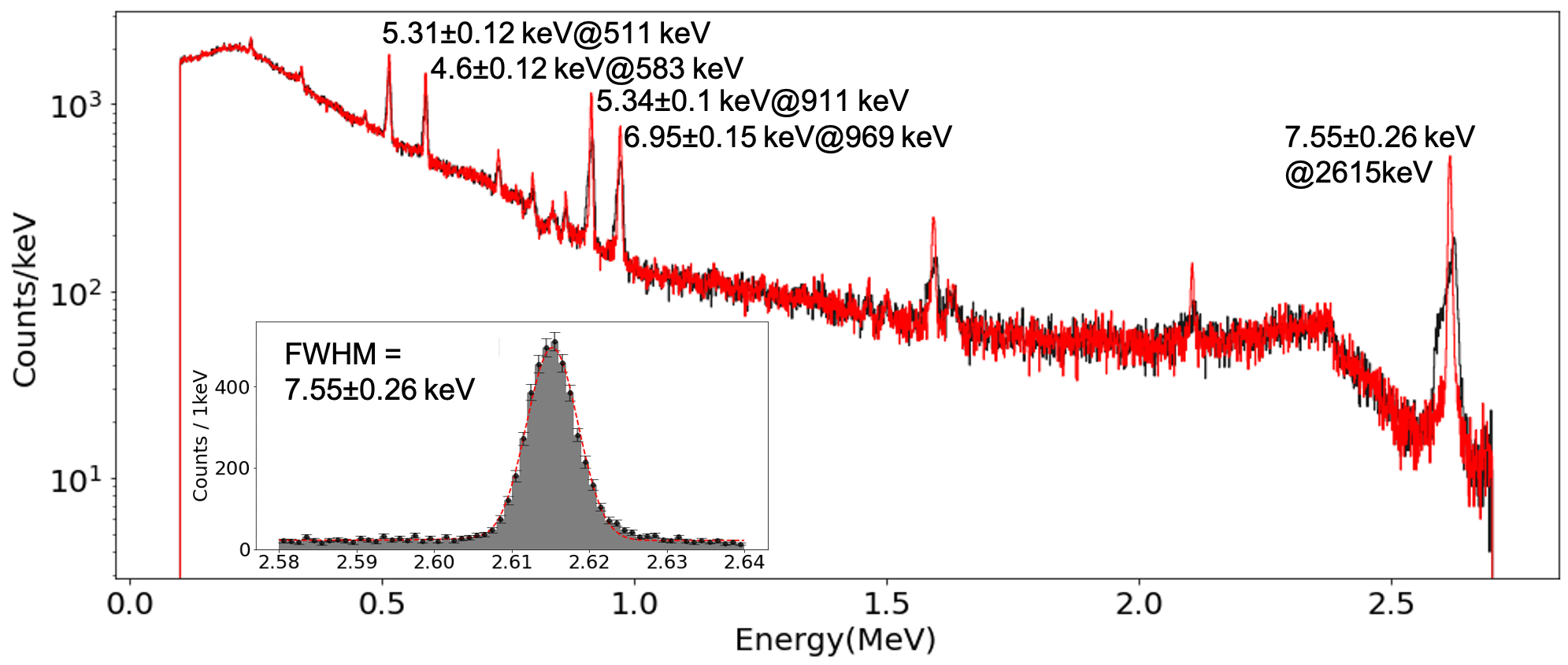}}
\caption{DC baseline dependent event energy for 2.615 MeV gamma rays with DET3 at 10mK.
Baseline versus event energy before (a) and after (b) the baseline dependent correction (``baseline correction")
(c) Energy spectrum
before (black) and after (red) the baseline correction. The energy resolution (FWHM) was estimated with a Crystal Ball function for
selected peaks with the corrected spectrum. The inset shows the 2.615 MeV peak (after correction) fitted with a Crystal Ball function.
}
\label{fig:baseline_vs_E}       
\end{figure}

The DC voltage signal from an MMC is proportional to the detector temperature, which reflects the transient temperature change due to the particle interaction in addition to tiny fluctuation of the baseline temperature. 
Therefore, it is possible to correct for temperature-dependent signal gain by analyzing the dependence of the height of pulses corresponding to a well defined energy with the average value of the DC pre-trigger samples (baseline) of the event pulses. 
The relation between the signal height and the baseline was investigated with the 2.615 MeV gamma line, as shown in Fig.~\ref{fig:baseline_vs_E}, and was used for the correction. 
The improvement in energy resolution across overall energy range
is clearly shown in Fig.~\ref{fig:baseline_vs_E}. 
Although the R\&D setups do not have stabilization heaters for drift correction, the baseline correction
effectively recovers the energy resolution from temperature drift. 
Previously, the spectral shape of the peaks in the measured energy spectrum were typically approximated as Gaussian shapes. However, there are non-Gaussian components, although subtle,  arising from various factors such as event position. For example, if events occur near the phonon collectors, their energy tends to be overestimated. 
Additionally, we sometimes observe a left-hand shoulder on the peak spectrum due to imperfect energy collection. 
For this reason, we adopted a Crystal Ball function~\cite{crystalball} to describe the shape of the energy peaks:

\begin{align}
f (x; \alpha,n,\bar{x}, \sigma) = N\cdot\begin{dcases*} 
 \text{exp}\left(-\frac{(x - \bar{x})^2}{2\sigma^2}\right)&for $\frac{x - \bar{x}}{\sigma} > - \alpha$\\
 A\cdot \left(B - \frac{x - \bar{x}}{\sigma}\right)^{-n}&for $\frac{x - \bar{x}}{\sigma} \leq - \alpha$
 \end{dcases*}
 \end{align} 
where \\
\begin{align}
&A  =  \left(\frac{n}{|\alpha|}\right)^n\cdot\text{exp}\left(-\frac{|\alpha|^2}{2}\right) \\
&B  = \frac{n}{|\alpha|} - |\alpha|
\end{align}
Here, $N$ is a normalization factor, and $\alpha$, $n$, $\bar{x}$ and $\sigma$  are fitting parameters.
In this report, we use the energy resolutions defined as FWHM of the Crystal Ball functions,
even though the values are very close to the FWHM with the Gaussian approximation with differences within a few percent. An illustration of a peak from 2.615 MeV gamma events fitted with a crystal ball function is shown 
in Fig.~\ref{fig:baseline_vs_E} (c).  Energy resolutions of various calibration peaks at 10 -- 30 mK temperatures are summarized in Table~\ref{tab:performance}. 
At the 2.615 MeV gamma line, the calibration peak close to the ROI,
the detectors showed FWHM of 7.55 -- 8.82 keV at 10 mK, 8.48 -- 12.34 keV at 20 mK and 10.74 -- 12.84 keV at 30 mK. 
The AMoRE-I detectors showed 10 -- 20 keV FWHM at the same calibration peak with the detector operation at 12 mK.
The overall energy resolution performance is significantly improved
compared to AMoRE-I experiments even though we used larger crystals than those in AMoRE-I and included operation at higher temperatures. 
In setting the target discovery sensitivity for AMoRE-II, we assumed energy resolution of 10 keV FWHM at the ROI. 
The results show that the energy resolutions sufficiently meet the design value of AMoRE-II
at 10 mK and remain acceptable even up to 30 mK operation.
This demonstrates that the detectors developed for AMoRE-II are not extremely sensitive to the operation temperature and 
allows for the operation of detectors at the higher temperature ($\sim$ 30 mK), where the pulse response is much faster implying better pile-up rejections with moderate sacrifice of the energy resolution. 
When operating AMoRE-II, the operating temperature can be tuned to optimize energy resolution and DP, which are better at lower temperatures, as well as timing resolution for pile-up rejections, which is better at higher temperatures.

\subsection{Light channel signal}
 
\begin{figure}
\centering
\subfloat[][\centering ]{\includegraphics[width=0.9\linewidth]{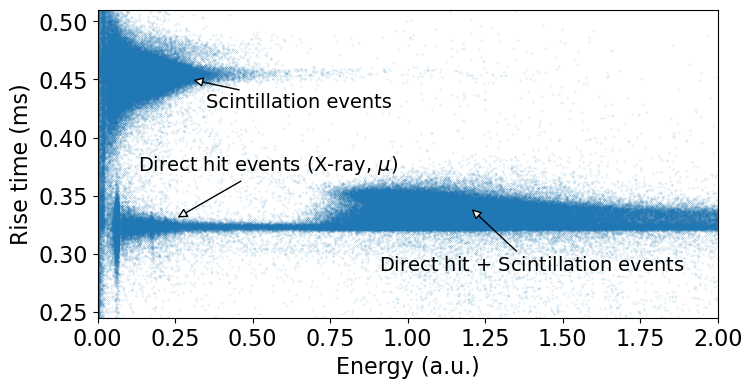}}\\
\subfloat[][\centering ]{\includegraphics[width=0.9\linewidth]{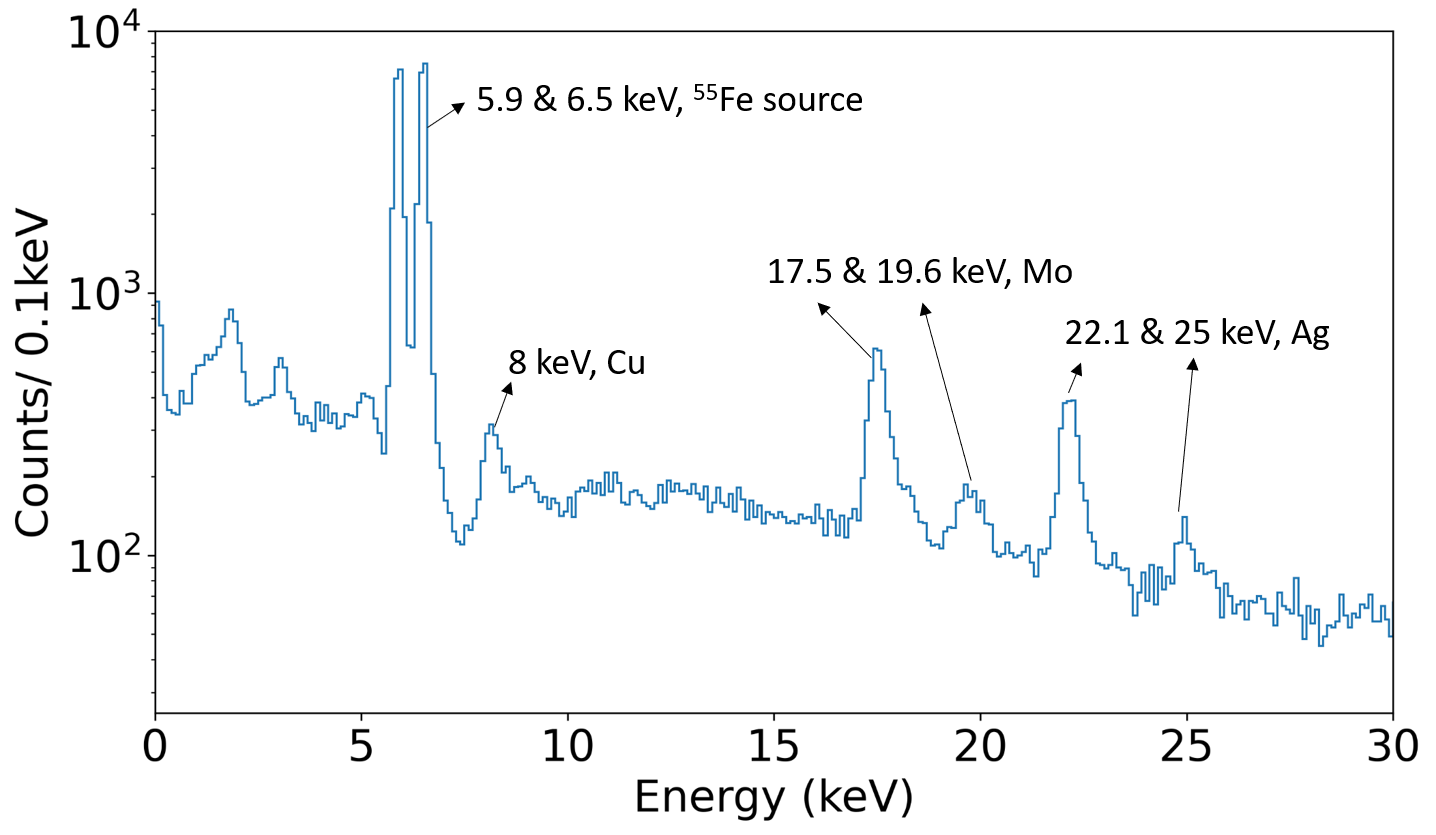}}\\
\subfloat[][\centering ]{\includegraphics[width=0.9\linewidth]{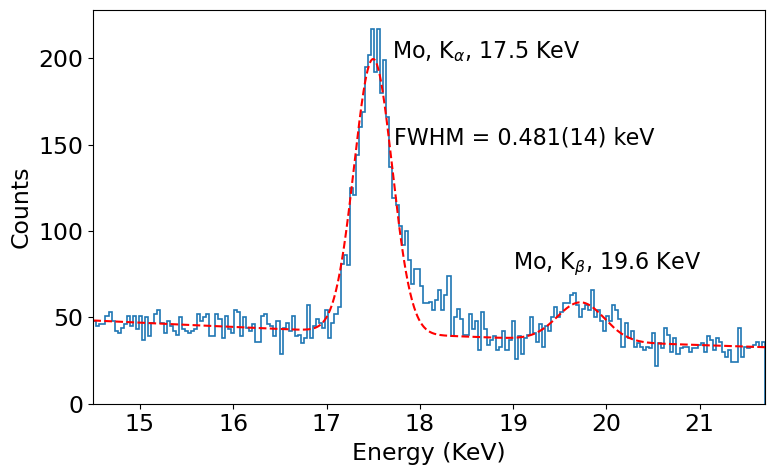}}
\caption{Light channel signal analysis with DET2 at 10 mK. (a) Rise time versus energy. (b) The energy spectrum of the direct hit events such as X-rays and muons. (c) Energy peak at X-rays from Mo nuclide fitted with 
a Gaussian shape. }\label{light_ana}
\end{figure}

The MMC-based light detectors were developed to be integrated in the AMoRE modules~\cite{hjlee_LD2015,mbkim_ieee2023}.
Although these light detectors themselves demonstrated high sensitivities, 
the light channels in previous AMoRE and R\&D setups were found to be more vulnerable to vibrations from the operation of a pulse tube refrigerator (PTR). 
The energy input from the absorption of scintillation light is small and easily swamped by the low-frequency noise (a few Hz) associated with cryostat vibration. 
In the pulse-height estimation for the light channels, bandpass filters were applied similarly to the heat channel~\cite{wtkim_jinst2022,wtkim_nmo_2022}.
We set the passband frequencies as 400 -- 900 Hz for the light channel signals. 
Figure~\ref{light_ana} (a) shows an example plot of the rise times of the signals versus their energies, indicating two distinct groups in the rise time distribution. 
The signals from the measurement of scintillation light show
greater rise time values centered near 450 $\mu$s, while other non-scintillation
events, such as X-rays and muons, referred to as direct hit events in this report, indicate
shorter rise times. 
This difference originates from the scintillation decay time, on
the order of 100 $\mu$s, appearing in LMO crystals~\cite{bagrat_ieee2023,bijaya_jltp2023}.
The energy of the scintillation signals can be calibrated with the direct hit X-ray signals.
As demonstrated in Fig.~\ref{light_ana} (b) and (c), 
this detector obtained 198 eV FWHM for 5.9 keV X-ray signals from an $^{55}$Fe source covered by aluminum foils with a total thickness of a few hundreds of micrometers, 
irradiating a large area of the wafer surfaces, and 481 eV FWHM for 17.5 keV X-rays
from Mo in the crystal. 
We assumed a Gaussian shape for the energy peaks in the light channel signals. 
The energy resolutions at baseline and the 17.5 keV X-ray peak for the overall R\&D detectors are summarized in Table~\ref{tab:performance}.
The baseline energy resolutions were measured as 83 -- 383 eV at 10 mK, 101 -- 640 eV at 20 mK and 159 -- 775 eV at 30 mK.
At the 17.5 keV peak, the estimated energy resolution was 481 - 850 eV at 10 mK, 573 -- 1190 eV at 20 mK and 750 -- 1290 eV at 30 mK. 
Overall, the R\&D detectors showed a modest decline of the energy resolution as the operation temperature increased, yet maintained good performance up to 30 mK.  
 The energy resolution of the light detectors at other X-ray peaks is fully discussed in other report~\cite{bijaya_jltp2023}. 

We performed a convolution fit to estimate the energy of scintillation light actually detected, assuming two-component exponential decay for the scintillation of the LMO crystals~\cite{bijaya_jltp2023}.  At 10 mK, the fraction of deposited energy observed as scintillation light was obtained as 0.79 - 0.96 keV/MeV using the events of 2.615 MeV gamma rays as the reference signal (Tab.~\ref{tab:performance}). This scintillation detection remains almost the same up to 30 mK. The quenching of the scintillation light for 
alpha particles is essential because it allows the discrimination of alpha particles based on the light/heat signal ratios. We compared the averaged pulses for 2.615 MeV gamma rays
and 4.785 MeV alpha particles with tritium from the neutron capture on $^6$Li, which can be considered as alpha events, and estimated the ratio of these pulse heights after normalizing each pulse by its own energy. 
We found that the quenching 
for alpha signals is around 0.25 for every detector in the tested temperature range, as shown in Table~\ref{tab:performance}. It means that the amount of scintillation energy collected from alpha particles is about 25\% as much as that collected for full-energy deposition of gamma rays with the same energy as the alpha particles. 

\begin{figure}
\centering
\includegraphics[width=0.95\linewidth]{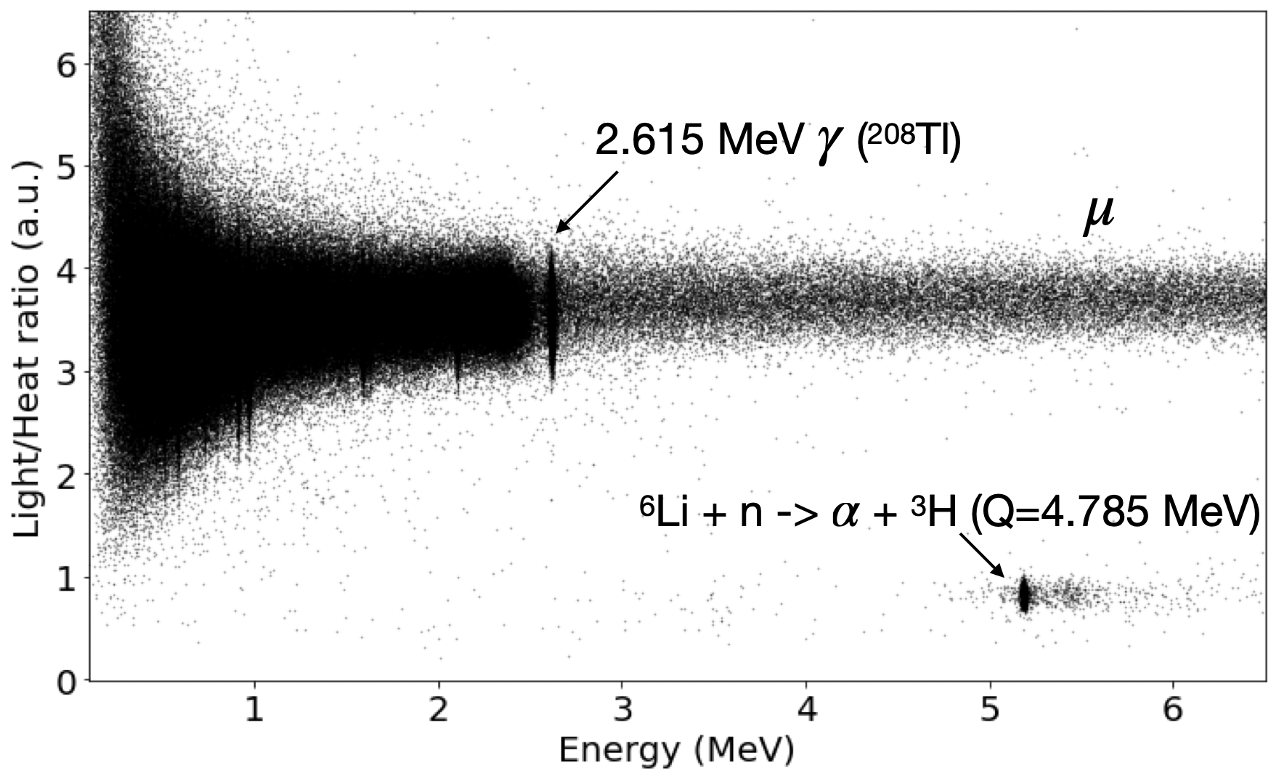}
\caption{Light/heat signal ratio versus energy of the events. This plot is for DET2 at 10 mK.}\label{ltoh}
\end{figure}

Figure~\ref{ltoh} shows the light/heat signal ratio versus the energy of the signals measured with DET2 at 10 mK.
Two groups of events are notable. A band of events with greater light/heat ratios originates from beta, gamma rays and muons. The light/heat ratio of $^6$Li neutron capture events is, indeed, approximately 25\% of that observed in the band with beta, gamma, and muon events.
This plot indicates that degraded alphas, which release only a fraction of its energy
in the crystal due to their occurrences at the surface, can also be well discriminated. 
We estimated DP between the muon and the alpha signals around the signals from the $^6$Li neutron capture to be 12.37 -- 19.50 at 10 mK as shown in Table ~\ref{tab:performance}. Interestingly, the DP based on the light/heat ratio at 30 mK 
is as high as 7.04 - 15.78, which still can allow excellent alpha signal rejections.
We tried to project the DP value at the ROI (E = 3.034 MeV) based on the measured scintillation detection (number of photons for a given energy). 
We introduced an additional energy-independent spread in the scintillation detection to account for the measured DP around E = 4.785 MeV. 
The projected values (L/H (ROI)) are also presented in Table 3. 
The projected DP ranges from 8.1 to 13.1 at 10 mK, reaching 4.5 -- 10.6 at 30 mK.       
The improved DP values compared to the AMoRE-I experiment are mainly attributed to
the detector design.
The AMoRE-II setup features a small gap between the light absorber and the crystal, along with more efficient coverage by the reflector film, increasing light collection efficiency~\cite{cupid_detectors_epjc2021}. Additionally, the setup introduced a tightened  grip on the light absorber in the detector holder, reducing low-frequency noise caused by cryostat vibration.  

To illustrate the time response of the scintillation light signals, 
the averaged signal for 2.615 MeV gamma rays with DET2 is shown in Fig.~\ref{signal_template} (b).
The characteristics of the scintillation signal for all four detectors are summarized in Table.~\ref{tab:pulses}. The rise time of the scintillation signals is 461 -- 873 $\mu$s at 10 mK, 403 -- 721 $\mu$s at 20 mK
and 373 -- 582 $\mu$s at 30 mK. The faster time responses, in comparison to the heat channel signals, can be exploited to enhance the pile-up rejection capability and may allow for bigger crystal absorbers,
such as crystals of 7 cm (D) $\times$ 7 cm (H) dimensions, which are already grown by NIIC to be tested for the target absorber. 
Except for detector DET4, the decay time of the scintillation signals are in the range of 10 -- 30 ms,
with faster response at the higher temperatures. DET4 shows
a very slow decay compared to other detectors. We suspect that the thermal link on the MMC was somehow damaged.

\section{Conclusions}

We successfuly tested the detector modules designed for AMoRE-II.
The thermal calorimetric detection for both heat and light signals from lithium molybdate crystals was carried out using MMC sensors. 
During the study to optimize detector performance, we established the protocols
for crystal handling, including surface treatment, cleaning, packaging, and storage.
The AMoRE collaboration has invested considerable effort in establishing reliable production
lines for numerous massive crystals with low radioactive backgrounds. As a result,
high-quality crystals can now be produced both at CUP and NIIC. 
We built R\&D setups using two kinds of massive cylindrical crystals with  dimensions of 5.1 cm (D) $\times$ 5 cm (H) and 6 cm (D) $\times$ 6 cm (H), featuring diffusive crystal surfaces and new detector holder for AMoRE-II.  
The R\&D tests demonstrated 
good energy resolutions for both heat and light channels signals,
as well as high discrimination power to reject alpha particles in the temperature range between 10 and 30 mK.
The pulse timing of the heat channel was shown to be reasonably good for the pile-up rejection.
In designing the AMoRE-II experiment, the benchmark values for the detector performance are an energy resolution of 10 keV FWHM and DP of 5 at the ROI, and a rise time of 5 ms for the event pulses. The characterization of the R\&D detectors indicates that they operate in a promising manner within the tested temperature range of 10 -- 30 mK. 
This detector performance demonstrated in this report will
serve as a baseline milestones toward the large-scale
AMoRE-II experiment, which will probe the nature of neutrinos with significantly improved sensitivity.

\afterpage{
\onecolumn
\begin{landscape}
\centering
\begin{table*}
\caption{Summary of the R\&D detector performances at 10, 20, and 30 mK operation temperatures.
The energy resolutions are presented at 583 keV ($^{208}$Tl), 911 keV ($^{228}$Ac), and 2615 keV ($^{208}$Tl) for the heat channel
and 17.5 keV (Mo X-ray) for the light channel in addition to the baseline. 
The energy peaks are modeled as Crystal Ball functions for the heat channel signals and Gaussian functions for the light channel signals. The energy resolutions are defined as the FWHM of each modeling function shape.
The light collection is estimated with the scintillation of 2.615 MeV gamma rays.
The light quenching is the ratio of light collection by alpha signals to that of gamma signals assuming the same energy. 
Discrimination powers based on PSD and light/heat signal (L/H) ratio are presented in the last columns.
These values are estimated using the muons and alphas within a $\pm$ 0.5 MeV energy window around $^6$Li neutron capture signals (E = 4.785 MeV).
DPs for PSD are presented when the separations between the muon and alpha signals are well-defined. The projected values for DP with L/H at the ROI are shown in the column of L/H (ROI).  
}
\label{tab:performance}       
\centering
\begin{tabular}{>{\centering\arraybackslash}m{1.2cm}
>{\centering\arraybackslash}m{1.8cm}
>{\centering\arraybackslash}m{1.2cm}
>{\centering\arraybackslash}m{1.2cm}
>{\centering\arraybackslash}m{1.3cm}
>{\centering\arraybackslash}m{1.3cm}
>{\centering\arraybackslash}m{0cm}
>{\centering\arraybackslash}m{1.3cm}
>{\centering\arraybackslash}m{1.3cm}
>{\centering\arraybackslash}m{1.4cm}
>{\centering\arraybackslash}m{1.5cm}
>{\centering\arraybackslash}m{1.1cm}
>{\centering\arraybackslash}m{1.1cm}
>{\centering\arraybackslash}m{1.6cm}
}
\hline\noalign{\smallskip}
\multirow{2}{*}{\adjustbox{stack=cc}{Detector name}} & \multirow{2}{*}{\adjustbox{stack=cc}{Operation temperature (mK)}} 
&\multicolumn{4}{c}{\parbox{3.4cm}{Heat channel \\ energy resolution (keV)}} 
&&\multicolumn{2}{c}{\parbox{3.4cm}{Light channel \\ energy resolution (eV)}} 
& \multirow{2}{*}{\adjustbox{stack=cc}{Light collection (keV/MeV)}} & \multirow{2}{*}{\adjustbox{stack=cc}{Light quenching for $\alpha$}}
& \multicolumn{3}{c}{Discrimination power }
\\
\noalign{\smallskip}
\noalign{\smallskip}
\cline{3-6} \cline{8-9} \cline{12-14}
& & Baseline & 583 keV & 911 keV & 2615 keV && Baseline  & 17.5 keV & &  & PSD & L/H & L/H (ROI)\\
\noalign{\smallskip}\hline
\multirow{3}{*}{\adjustbox{stack=cc}{DET1}} & 10 & 2.57(2) & 4.58(15) & 5.33(12) & 8.55(22)   && 173(3) & 850(60)  & 0.87(2) & 0.252 & - & 12.37(9) & 8.1\\
& 20 & 3.27(10)  & 4.58(13) & 4.91(13) & 8.48(21)  && 202(3)  & 820(60)  & 0.78(2)  &0.257 & - & 11.20(12) & 7.3\\
& 30 & 7.58(2)  & 7.88(31)  & 7.87(22)  & 10.74(26)  && 342(5)  & 850(90)  & 0.72(3)  &0.274 & - & 8.84(23) & 5.7\\
\hline
\multirow{3}{*}{\adjustbox{stack=cc}{DET2}} & 10  & 3.34(2) & 4.74(12)  & 5.37(11)  &  8.35(19)  && 99(1) & 481(14)  & 0.83(4) & 0.250 & 3.41(8) & 19.50(9) & 13.5 \\
& 20 & 3.54(8) & 4.46(11)  & 5.44(11) &  8.60(11) && 149(2) & 573(24) & 0.87(4) & 0.253 & 2.96(25)  & 14.64(11) & 9.7\\
& 30 & 6.25(10)  & 8.11(21) & 8.81(17)   &  11.74(20) && 351(6) & 750(40) &0.84(5) & 0.252 & - & 10.34(19) & 6.7 \\
\hline
\multirow{3}{*}{\adjustbox{stack=cc}{DET3}} &10 & 3.39(1) & 4.55(11)   & 5.43(10)  &  7.55(25) && 383(5) & 790(110) &0.96(15) & 0.255 & 3.25(20) & 13.82(16) & 8.8  \\
& 20 & 6.0(6) & 7.65(27) & 8.19(24) &  10.92(22) && 640(11) & 1190(140) & 0.95(18) & 0.257 & 2.03(10)  & 8.38(9) & 5.4  \\
& 30 & 8.95(21) & 9.27(25) & 9.40(18) &  12.84(20) && 775(15)  & 1290(150) &1.0(2) & 0.250 & - &  7.04(29) & 4.5 \\
\hline
\multirow{3}{*}{\adjustbox{stack=cc}{DET4}} & 10 &  3.20(2) & 5.85(18)   & 5.99(12)  &   8.82(16) &&83(1)  & 810(60) & 0.79(7) &0.259 & - &  18.89(8) & 13.1 \\
& 20 & 4.69(17) & 5.94(18) & 7.67(15)   &  12.34(23)  && 101(1) & 780(50) & 0.86(5) & 0.250 & - &  17.55(2) & 11.9 \\
& 30 &  8.27(9)  & 9.09(34)   & 9.31(24)   &  12.16(24)  && 159(2)   & 820(70) &0.91(7) & 0.254 & - &  15.78(11) & 10.6\\
\hline
\end{tabular}

\end{table*}

\end{landscape}
\twocolumn
}

\section{Acknowledgment}
This research was funded by the Institute for Basic Science (Korea) under
project codes IBS-R016-D1 and IBS-R016-A2. It is also supported by the 
National Research Foundation of Korea (NRF-2021R1I1A3041453, 
NRF-2021R1A\-2C1013761, NRF-2023K2A9A2A22091340), the National Research Facilities \& Equipment Center 
(NFEC) of Korea (No. 2019R1A6C1010027), and the MEPhI Program Priority 2030 of Russia.
The work at NIIC was supported by the Ministry of Science and Higher Education of the Russian Federation N1210317\-00314-5.
Loredana Gastaldo acknowledges support by the Deutsche Forschungsgemeinschaft through the contract GA 2219/4-1.
The group from the INR (Kyiv, Ukraine) was supported in part by the
National Research Foundation of Ukraine under Grant No. 2023.03/0213.
These acknowledgments are not to be interpreted as an endorsement of
any statement made by any of our institutes, funding agencies, governments,
or their representatives.

%
%

\end{document}